\documentstyle[prb,aps,amssymb,amsbsy]{revtex}  
\tightenlines
\begin{document}    
\draft
\title{	Spin filter effect of the europium chalcogenides -\\
		An exactly solved many-body model} 
\author{	R.~Metzke\thanks{e-mail:robert.metzke@physik.hu-berlin.de} and 
		W.~Nolting} 
\address{	Lehrstuhl Festk\"orpertheorie, Institut f\"ur Physik\\ 
		Humboldt-Universit\"at zu Berlin,
		Invalidenstra{\ss}e 110, D-10115 Berlin, Germany} 
\date{\today} 
\maketitle

\begin{abstract}                
A  model Hamiltonian
is introduced which considers the main features  of the experimental spin
filter situation as s-f interaction, planar geometry  and the strong external
electric field. The proposed many-body model can be  solved analytically and
exactly using Green functions. 

The spin polarization of the field-emitted electrons  is expressed in terms of
spin-flip probabilities, which on their part are put down to  the exactly known
dynamic quantities of the system.

The calculated electron spin polarization shows remarkable dependencies on  the
electron velocity perpendicular to the emitting plane and the strength of s-f
coupling. Experimentally observed polarization  values of about 90\% are well
understood within the framework of the proposed  model. 
\end{abstract}
\pacs{75.50.Pp,75.70.-i,75.40.Gb,73.40.Gk,}

\section{Introduction} 
\label{sec:intro} 
Because of their extraordinary
magnetic, optic and transport properties  magnetic semiconductors, among them
the europium chalcogenides, have been subject to numerous experimental and
theoretical investigations \cite{Wac79,No79a}. Many of them focused on the spin
filter effect of the europium chalcogenides (SFE).

The spin filter experiment can be arranged as a field emission experiment. 
Here a cooled tungsten emitter is exposed to a strong stationary electrical
field bending down the  vacuum level of the electric potential outside the
emitter. The so formed potential barrier  can quantum-mechanically be
penetrated by the conduction band electrons of the emitter metal. The
probability for this tunneling process is known to depend exponentially on the
barriers height. This strong dependence can be used to obtain a  ''spin
filter'' by covering the original emitter with a layer of a ferromagnetically
ordered material, e.g. a ferromagnetic semiconductor\cite{MEH72} as EuS
that makes the barrier spin dependent.

From an experimental point of view the generation of highly polarized electron
beams became more and more interesting with the growing importance of the spin
polarized electron spectroscopies which are nowadays a mighty tool in  the
field of analyzing magnetic properties of surfaces and thin films 
\cite{Do93,SchoSie93,All94}.  The spin filter effect of the europium
chalcogenides (SFE) allows for polarization  values of about 90\%
\cite{MEH72,KiBM76,BKiM77,KiBM78}.

Examining the SFE  theoretically is worth while, too. Europium chalcogenides
(EuX, with X=O, S, Se, Te ) show highly interesting correlation effects due to
the  complex interplay of  itinerant conduction band electrons and  localized
4-f electrons, the latter  carrying a strong magnetic moment.  The so called
\emph{s-f model} \begin{eqnarray} \label{sfhamilton} H&=&H_s+H_{sf}\nonumber\\
&=&\sum_{ij\sigma} T_{ij} c_{i\sigma}^+c_{j\sigma}-\frac{J}{\hbar}\; \sum_{i} 
\boldsymbol{\sigma}_i \cdot \mathbf{S}_i  \end{eqnarray}  turns out to describe
the interaction  of both electron groups quite successfully\cite{Nol79a,RHB67}.
$c_{i\sigma}^{(+)}$ are  the usual annihilation- (creation-) operators for
conduction band electrons with spin $\sigma$ at site $\mathbf{R}_i$.   The spin
of a conduction band electron at site $\mathbf{R}_i$ is denoted by
$\boldsymbol{\sigma}_i$  and $\mathbf{S}_i$ represents the spin of the half
filled 4f-shell at this site.  (Notation is conventional.)

The simplest possible approximation to solve the corresponding many-body
problem is the mean field approximation of (\ref{sfhamilton}): \begin{eqnarray}
H^{MF}&=&\sum_{ij\sigma} T_{ij} c_{i\sigma}^+c_{j\sigma} -\frac{1}{2}J \langle
S^z\rangle\sum_{i,\sigma} z_{\sigma}n_{i\sigma}\nonumber\\ &=&\sum_{ij\sigma}
(T_{ij}-\frac{z_{\sigma}}{2}J \langle S^z\rangle\delta_{ij})
c_{i\sigma}^+c_{j\sigma} \end{eqnarray} 
($z_{\uparrow}=+1,\; z_{\downarrow}=-1$) that  has  often been  used to discuss
the spin filter experiment,  too\cite{KiBM76,BKiM77,KiBM78,Sig75}. The
mean-field decoupled s-f interaction term  spin dependently  renormalizes the
one-particle energies of the free conduction band electrons Fig.\ \ref{fig1},
giving them an explicit spin- and temperature dependence. 

Below $T_{c}$ the conduction band splits due to the interaction of conduction 
band and  magnetically ordering 4-f spin lattice into two  completely
spin-polarized sub-bands. This splitting is temperature dependent  and at $T=0$
of the order of $g\hbar S\approx1eV$. 

For the spin filter experiment this would mean, that below $T_c$ only 
$\sigma=\uparrow$-electrons could be emitted, as the sub-band for
$\sigma=\downarrow$-electrons lies much higher then the
$\sigma=\uparrow$-sub-band and $\downarrow$-electrons ''see'' accordingly a
much higher tunnel barrier.

The mean field picture thus predicts a degree of polarization for  the emitted
electrons that should be very close  to  100 \% for all  temperatures  below
$T_c$ and  0\% above.  However, this does not agree very well with the
experimental  data \cite{KiBM78}. Main failure of the mean field approximation
is probably the  complete suppression of spin-flip processes. 

The knowledge about the s-f model has multiplied 
\cite{No79a,BNoB87,NoDB93,NMJR96,quantmag2,Taka97b} since the  first attempts
of applying it to the SFE in the early 70-ties  \cite{Sig75,KiBM76,NoRei79}.
Aim of the present paper is a new interpretation of the SFE based on  meanwhile
made progress that permits us to treat the many-body problem of the spin filter
effect- even including the external electric field - exactly.

\section{Electron Spin Polarization} 
\label{sec:pol} 
\subsection{Polarization and Probabilities} 
In this section it will be shown, how the central quantity
of the spin filter experiment - the polarization - can be connected with the
dynamic quantities of the system,  the Green functions, which will  be
determined in the following section from the many-body model.

The vector \emph{Spin-Polarization} $\mathbf{P}$ of an ensemble of electrons
is  defined as expectation value of the Pauli spin-matrices. More easily to
handle is its projection on the preferential direction of spins, the scalar
quantity $P$ (\emph{degree of polarization}):   \begin{equation}\label{defp}
P\stackrel{\rm def}{=} 
\frac{N_{\uparrow}-N_{\downarrow}}{N_{\uparrow}+N_{\downarrow}}, \end{equation}
where $N_{\uparrow}$ is the number of electrons with spin $\sigma=\uparrow$,
$N_{\downarrow}$ the number of electrons with spin $\sigma=\downarrow$. We now
consider the probabilities for electrons with original spin $\sigma$ to flip
their spin, i.e. the \emph{flip-probabilities} $p_{\sigma}$. If for example the
originally prepared  state was $\downarrow$, then $p_{\downarrow}$ gives the
probability to measure an electron  in the $\uparrow$-state.\\ Assuming an
unpolarized beam of electrons ($N_{0\uparrow}=N_{0\downarrow}$) passing
through  a spin filter box we obtain  very easily the polarization of the
out-coming beam using (\ref{defp}) and the flip-probabilities $p_{\sigma}$
which are completely  determined by the physical properties of the spin filter
box: \begin{eqnarray} \label{nup}
N_{\uparrow}&=&N_{0\uparrow}(1-p_{\uparrow})+N_{0\downarrow}p_{\downarrow}=\frac{N_0}{2}
(1+p_{\downarrow}-p_{\uparrow}),\\ \label{ndown}
N_{\downarrow}&=&N_{0\downarrow}(1-p_{\downarrow})+N_{0\uparrow}p_{\uparrow}=\frac{N_0}{2}
(1-(p_{\downarrow}-p_{\uparrow}))\\ \label{p=w}
P&=&\frac{\frac{N_0}{2}\left(1+p_{eff}-(1-p_{eff})\right)}{N_0}=p_{eff}
\end{eqnarray} Here we introduced $\quad\displaystyle p_{eff}\stackrel{\rm
def}{=} p_{\downarrow}-p_{\uparrow}$: the effective flip-ratio of the spin
filter.

Asking for the polarization, we should thus try to get the flip-probabilities
from a theoretical model, i.e. Green functions.

\subsection{Probabilities and Green Functions} 
\label{subsec:prob_green} 
Let us
consider the following example:\\  What is the (non flip) probability
$\bar{p}(t)$ to measure at time $t$ an  electron with  wave-vector
$\mathbf{k}$, if at  $t=0$ an electron in this state had been prepared?\\  

The answer is given\cite{Born26b} by the overlap of initial state
$c_{\mathbf{k}}^+(0)|0\rangle$ and final state $c_{\mathbf{k}}^+(t)|0\rangle$.
$c_{\mathbf{k}}^{(+)}(t)$ (creates) annihilates an electron at time $t$ with
wave-vector $\mathbf{k}$. The state $|0\rangle$ is the electron- and magnon-
vacuum. \begin{equation}\label{w=ckck} \bar{p}(t)=\big| \langle
0|c_{\mathbf{k}}(t)\;\; c_{\mathbf{k}}^+(0)|0\rangle \big|^2\;. \end{equation}
One sees the similarity of the probability $\bar{p}(t)$ with a well known 
entity of many-body theory, the time-dependent spectral density of the
one-electron Green function: \begin{equation}\label{defS} S_{\mathbf{k}}(t,0)=
\frac{1}{2\pi}\langle[c_{\mathbf{k}}(t),c_{\mathbf{k}}^+(0)]_+\rangle
\end{equation} In our special case (T=0, n=0),  the average denoted by the
$\langle\rangle$-brackets has  to be taken with the $|0\rangle$-state, i.e.
with the electron- and magnon-  vacuum. Therefor one of the terms in the
anticommutator does not contribute.

Spectral density (\ref{defS}) and Green function 
$G_{\mathbf{k}\sigma}(E)\equiv\left\langle\!\left\langle
c_{\mathbf{k}\sigma};c_{\mathbf{k}\sigma}^+\right\rangle\!\right\rangle_E$ are 
closely related. Both will be determined in section (\ref{sec:sol}).\\ From
(\ref{w=ckck}) and (\ref{defS}) we obtain immediately 
\begin{equation}\label{w=S1} \bar{p}(t)=4\pi^2 \;\big| S_{\mathbf{k}}(t)\big|^2
\;. \end{equation}

\section{Theoretical Model} 
\label{sec:model}

Aim of this section is to develop a many-body model of the spin filter which
incorporates the main features of the experimental situation.

We make the following two assumptions, which allow us to treat the 
corresponding many-body problem exactly:  \begin{enumerate} \item The tunneling
processes of the emitted electrons are  independent from each other; i.e.  the
conduction band is nearly  empty (n=0). \item The 4f-lattice is
ferromagnetically saturated, i.e. low temperatures (T=0). \end{enumerate}

\subsection{Geometry}

Our calculations are done for  a \emph{planar}  spin filter, i.e. a
sandwich-like batch of mono layers, which are translational invariant parallel
to their plane. This matches  the experimental situation (field emission
through a EuS-layer) and provides at the same time the proper symmetry  to
include the influence of the strong electric field.

The treatment of a planar system  is based on the decomposition of the whole
system into $n$ equivalent  two-dimensional sub-lattices (atomic layers) with
$N_{s}$ lattice points each: \begin{equation}\label{zerl}
\mathbf{R}_{i\alpha}=\mathbf{R}_{i}+\mathbf{r}_{\alpha}. \end{equation}
$\mathbf{R}_{i}$ and $\mathbf{r}_{\alpha}$ are perpendicular to each other.  
$\mathbf{r}_{\alpha}$  points to the atomic layer with index $\alpha$, whereas
$\mathbf{R}_{i}$ points towards the lattice point with index $i$  inside this
layer.

From the lattice vector $\mathbf{R}_{i\alpha}$ this subscription is carried on
to \emph{all}  operators and derived quantities; greek indices generally
referring  to the layer,  latin indices to lattice points within this layer:
\begin{equation}\label{indizes} O_{i}\longrightarrow O_{i\alpha}, \qquad
O_{ij}\longrightarrow O_{ij}^{\alpha\beta},  \qquad O_{ikj}\longrightarrow
O_{ikj}^{\alpha\beta\gamma}. \end{equation}

Fourier transformation into $\mathbf{k}$-space is reasonably defined only
within the  two-dimensional sub-lattices, which are still invariant under
translation: \begin{eqnarray} \label{defftsch}
O_{i\alpha}&=&\frac{1}{\sqrt{N_{s}}}\sum_{\mathbf{k}} 
e^{i\mathbf{k}\mathbf{R}_{i}} O_{\mathbf{k}\alpha}, \\ \label{defftsch2}
O_{\mathbf{k}\alpha}&=&\frac{1}{\sqrt{N_{s}}}\sum_{i} 
e^{-i\mathbf{k}\mathbf{R}_{i}} O_{i\alpha}. \end{eqnarray} $\mathbf{k}$ means
through all following considerations a vector of the  two-dimensional 
Brillouin-zone (BZ) of one layer.

\subsection{s-f interaction}

It is convenient to write (\ref{sfhamilton}) in a for our purposes more 
suitable form: \begin{equation}\label{sfhamsch} H_{sf}=- \frac{1}{2} \,J
\sum_{i\alpha\sigma}(z_{\sigma} S_{i\alpha}^z n_{i\alpha\sigma}
+S_{i\alpha}^{\sigma}c_{i\alpha-\sigma}^+c_{i\alpha\sigma}). \end{equation}
Here the spin-operators for the conduction electrons  were expressed in terms
of the   creation- and annihilation- operators $c_{i\alpha\sigma}^{(+)}$:
\begin{eqnarray} \label{hsftrafo}
\frac{1}{\hbar}\,\sigma_{i\alpha}^+=c_{i\alpha \uparrow}^+c_{i\alpha \downarrow
},&\quad& \frac{1}{\hbar}\,\sigma_{i\alpha}^-=c_{i\alpha
\downarrow}^+c_{i\alpha \uparrow}\\
\frac{1}{\hbar}\,\sigma_{i\alpha}^z&=&\frac{1}{2} \sum_{\sigma}
z_{\sigma}n_{i\alpha\sigma}, \end{eqnarray}  with $z_{\uparrow}=+1,
z_{\downarrow}=-1$ and  the  identity  for the ladder operators
\begin{equation}\label{abk} S_{i\alpha}^{\uparrow(\downarrow)}\equiv
S_{i\alpha}^\pm = S_{i\alpha}^x  \pm iS_{i\alpha}^y \end{equation}  has been
used.

This representation already indicates some physics: The first term in
(\ref{sfhamsch}) describes the interaction of  the z-components of the spins
similar to the well known  Ising-model\cite{Ising25}.  Accordingly it will be
called \emph{Ising-term}. The second term in (\ref{sfhamsch}) is non-diagonal
in spin-indices and thus responsible for spin-flip-processes; it will be
referred to as \emph{spin-flip term}.

Interactions among the strongly localized 4-f spins might be  taken into
account via a Heisenberg-term \cite{Hei28}, but will here be neglected,  as we
are especially interested in the situation of conduction band electrons. Magnon
energies are two to three orders of magnitude smaller than the other energy
scales of the system as bandwidth or s-f coupling.

For similar reasons the interaction among different conduction band  electrons
will not be part of the model:  europium chalcogenides are magnetic 
semiconductors, ordering at fairly low temperatures (\mbox{$T_c^{\rm
EuS}=16.57$ K \cite{Wac79}).}  In  the interesting range of temperatures, the
conduction band is nearly empty, and thus the contribution of
electron-electron-interaction to the total energy of the system will be 
negligible small.

\subsection{External Electric Field}

The Hamiltonian of an external electric field  \begin{equation}\label{vhat}
H_V= - \mathbf{\widehat{P} \cdot F}  \end{equation} is a one-particle-operator.
Rewriting it in terms of second quantization: \begin{eqnarray} \label{1TOp}
H_{V}&=&\sum_n h_{V}^{(n)}\nonumber\\ &=&\sum_{ij\sigma \atop \alpha\beta}
\;M_{ij}^{\alpha\beta}\;c_{i\alpha\sigma}^+c_{j\beta\sigma}, \end{eqnarray} we
have to determine the matrix elements $M_{ij}^{\alpha\beta}$  .  $h_{V}^{(n)}$
acts in the one-particle-Hilbert-space of the  $n$-th electron;
$|\,i\alpha\,\rangle$ are elements of a complete orthonormal basis, for
instance \emph{Wannier}-states.

Calculating the $M_{ij}^{\alpha\beta}$, we have to specify the vector of
electric field $\mathbf{F}$, and  the operator of the electric  dipole momentum
$\mathbf{\widehat{P}}$.  The electric field is assumed to act homogeneously
along the z-axis, i.e. perpendicular to the EuX-film (see Fig.\ 
\ref{fig3}): \begin{equation}\label{feld} \mathbf{F}=\left(F_{x},
F_{x},F_{z}\right)=(0,0,-f=const.). \end{equation} With
\begin{equation}\label{dipol1} \mathbf{\widehat{P}}=-e \sum_n
\;\mathbf{\widehat{r}}^{(n)}\;, \end{equation}

we evaluate the matrix elements:  \begin{eqnarray} \label{mij}
M_{ij}^{\alpha\beta} &=& -ef\;\langle j\beta|  (\mathbf{\widehat{r}})_{z} 
|i\alpha\rangle\nonumber\\ &\cong& -ef
(\mathbf{R}_{i\alpha})_{z}\;\delta_{ij}^{\alpha\beta}\nonumber\\ &=& -ef\alpha
a_0 \;\delta_{ij}^{\alpha\beta}=-\alpha\,u\;\delta_{ij}^{\alpha\beta}.
\end{eqnarray} Here we assumed the Wannier functions to be eigen-functions of
the position-space operator, what should hold in good approximation.
Additionally we used
(${\mathbf{R}}_{i\alpha})_{z}\equiv|\mathbf{r}_{\alpha}|=\alpha a_0 $:  the
z-component of  lattice vector $\mathbf{R}_{i\alpha}$ equals the  product of
layer index $\alpha$ and lattice constant $a_0$.  Furthermore we introduced the
interlayer potential  difference $u=efa_0$.

For the Hamiltonian of a homogeneous field, acting along the z-axis, i.e. 
perpendicular to spin filtering EuX-film we eventually obtain
\begin{equation}\label{feldham} H_{V} = -u\; \sum_{i\alpha\sigma}
\;\alpha\,n_{i\alpha\sigma} \;. \end{equation} Finally we thus propose the
following model Hamiltonian for the spin filter:
\begin{equation}\label{hamiltonsch} H=H_{s}+H_{sf}+H_{V}. \end{equation}

\section{Solution of the Many-Body Problem} 
\label{sec:sol} 
\subsection{s-f model for  $T=0$, $n=0$}

The theoretical model developed in the previous section will now be solved. We
consider a single test-electron in the otherwise empty conduction band and a
ferromagnetic  saturated 4f-spin system. This special case of the s-f model  is
not only fundamental for the understanding of the spin filter effect  as it
meets the experimental conditions;  it moreover can be solved exactly
\cite{NoltingBd7,quantmag2}, even for the planar geometry \cite{SMN96}.

The retarded one electron anti-commutator-Green-Function
\begin{equation}\label{eegfschsf} G^{\alpha \beta}_{ij \sigma} (E) =
\left\langle\!\left\langle c_{i \alpha \sigma} ; c^+_{j \beta \sigma}
\right\rangle\!\right\rangle_E. \end{equation} will give us among other
interesting information the time-dependent spectral density, which is needed 
for calculating the flip-probabilities (\ref{w=S1}) and thus the polarization
of the emitted electrons (\ref{p=w}): 
\begin{eqnarray}
S_{\mathbf{k}\sigma}^{\alpha\beta}(t)&=&\frac{1}{2\pi\hbar}
\int_{-\infty}^{+\infty}dE\;  e^{-\frac{i}
{\hbar}E\,t}\;S_{\mathbf{k}\sigma}^{\alpha\beta}(E) \\
S_{\mathbf{k}\sigma}^{\alpha\beta}(E)&=&-\frac{1}{\pi}\;
\Im\{G_{\mathbf{k}\sigma}^{\alpha\beta}(E+i0^+)\}
\end{eqnarray}

To determine this Green function, one has  to solve its equation of motion:
\begin{equation}\label{algbwglsch} E G^{\alpha \beta}_{ij \sigma} (E) = \hbar 
\left\langle [c_{i \alpha \sigma} , c^+_{j \beta \sigma}]_+ \right\rangle +
\left\langle\!\left\langle [c_{i \alpha \sigma} , H]_{-} ; c^+_{j \beta \sigma}
\right\rangle\!\right\rangle_E, \end{equation} The higher Green functions,
appearing on the rhs of (\ref{algbwglsch}) \begin{eqnarray} \label{isingfkt}
\Gamma_{ik,j\sigma}^{\alpha\gamma \beta}(E)&=&\left\langle\!\left\langle
S^{z}_{i\alpha}c_{k\gamma\sigma};c_{j\beta\sigma}^+
\right\rangle\!\right\rangle_E, \\ \label{flipfkt} F_{ik,j\sigma}^{\alpha\gamma
\beta}(E))&=&\left\langle\!\left\langle
S^{-\sigma}_{i\alpha}c_{k\gamma-\sigma};c_{j\beta\sigma}^+
\right\rangle\!\right\rangle_E \end{eqnarray} are to be calculated by solving
\emph{their} equations of motion.   This procedure  usually leads  to an
infinite hierarchy of coupled Green functions and  their equations of motion.
To get any solution, physically reasonable decouplings are needed. One possible
treatment would be  the  mean field approximation  where (\ref{isingfkt}) can
be expressed in terms of (\ref{eegfschsf}) and (\ref{flipfkt}) is suppressed
completely. 

After  doing the two-dimensional Fourier transformation parallel to the planes,
the equation  of motion (\ref{algbwglsch}) of the one-electron Green function 
$G^{\alpha \beta}_{ij \sigma} (E)$ reads as follows: \begin{eqnarray}
\label{eom3} \sum_{\delta}\Big((E + \textstyle \frac{1}{2} \displaystyle
z_{\sigma} J \hbar S)\delta_{\alpha\delta}
&-&\epsilon_{\alpha\delta}(\mathbf{k})\Big) G^{\delta \beta}_{\mathbf{k}
\sigma}= \nonumber\\  &=&\hbar \delta_{\alpha \beta}\,  - {\small \frac{J}{2
\sqrt{N_s}} \sum_{\mathbf{q}} } F^{\alpha \alpha \beta}_{\mathbf{kq} \sigma}
\end{eqnarray}

In the  case of an empty conduction band (n=0) and ferromagnetic saturation of
the f-spin lattice (T=0) which is considered here,  we can solve this exactly.
Writing  down the equation of motion of  the so called flip-function 
$F_{ik,j\sigma}(E))$ (\ref{flipfkt}) one recognizes, that all higher Green
functions may be expressed in terms of already known ones ore  vanish
\cite{Nol79b,SMN96}. We find the following expression:
\begin{equation}\label{sumf=g} \frac{1}{\sqrt{N_s}} \sum_{\mathbf{q}} F^{\alpha
\alpha \beta}_{\mathbf{kq} \downarrow} = - \frac{J \hbar^2 S B_{\alpha}(E)}{1-
\frac{1}{2} J \hbar B_{\alpha}(E)} G^{\alpha \beta}_{\mathbf{k} \downarrow},
\end{equation} 
that expresses the flip-function completely in terms of the
one-electron Green function $G^{\alpha \beta}_{\mathbf{k}\sigma} (E)$. Here we
introduced the complex propagator $B_{\alpha}(E)$: 
\begin{equation}\label{b_al}
B_{\alpha}(E) = \frac{1}{N_s} \sum_{\mathbf{q}}\; G_{k-q\uparrow}^{\alpha
\alpha}(E)\;, 
\end{equation} 
(\ref{sumf=g}) and (\ref{eom3}) form a coupled system
of equations.  The equation of motion (\ref{eom3}) is at this point obviously 
equivalent to a matrix-multiplication which may be written in the following
compact formulation:
\begin{equation}\label{bwglmatrix}
\left(\mathbf{E}-\mathbf{H}_{\mathbf{k}\sigma}\right)\cdot\mathbf{G}_{\mathbf{k}\sigma}=\hbar\,\mathbf{I}.
\end{equation} $\mathbf{H_{k\sigma}}$ is the  effective  Hamiltonian of the
one-dimensional problem of an atomic chain perpendicular to the  translational
invariant layers.  In site representation it reads 
\begin{equation}\label{hk}
\mathbf{H}_{\mathbf{k}\sigma} = \left( \begin{array}{ccc}  
\begin{array}{lll}     a_1 & b_{12} & b_{13}   \\  b_{21} & a_2 & b_{23} \\
b_{31} & b_{32} & a_3 \end{array} & \cdots &  \begin{array}{l} b_{1n} \\ b_{2n}
\\ b_{3n} \end{array} \\ \vdots & \ddots & \vdots \\  \begin{array}{lll} b_{n1}
& b_{n2} & b_{n3} \end{array} &  \cdots & a_n \end{array} \right)
\end{equation} 
with its matrix elements: 
\begin{eqnarray} \label{mefsfup}
a_{\alpha} &=& \epsilon_{\alpha\alpha}(\mathbf{k}) + \Sigma_{\sigma}^{\alpha}(E)\\
b_{\alpha \beta} &=& \epsilon_{\alpha \beta} (\mathbf{k}) 
\end{eqnarray}
containing the complex, local and spin-depending  self-energy
$\Sigma_{\sigma}^{\alpha}$ 
\begin{equation}\textstyle
\Sigma_{\sigma}^{\alpha}(E)=-z_{\sigma}\,\frac{1}{2}\, J\hbar S\left(1+{
\frac{1-z_{\sigma}}{2}}\; \frac{J\hbar B_{\alpha}(E)}{1-\textstyle \frac{1}{2}
\displaystyle J\hbar B_{\alpha}(E)}\right)\;, 
\end{equation} 
the complex propagator $B_{\alpha}(E)$ (\ref{b_al}) and the Fourier-transformed
hopping-integrals  $\epsilon_{\alpha \beta} (\mathbf{k})$. It should be
noticed, that the self-energy for  $\sigma=\uparrow$-electrons is trivial. This
is due to the fact, that these electrons cannot participate in the spin flip
processes, so that their interaction with the ferromagnetically saturated 4-f
spins is restricted to a rigid energy shift of the
$\sigma=\uparrow$-conduction band. 

The propagator $B_{\alpha}(E)$ and the self-energy $\Sigma_{\sigma}^{\alpha}$ are
layer dependent  (index $\alpha$) but independent of the in-plane wave vector
$\mathbf{k}$. The latter is due to our neglect of magnon energies which is not
necessary but  convenient and reasonable since they are two to three orders of
magnitudes smaller than  other typical energy scales of the system (bandwidth,
s-f coupling). However, our model might be solved exactly with full
consideration of the magnon energies.

From (\ref{bwglmatrix}) we get finally the exact solution of the
many-body-system by matrix-inversion: \begin{equation}\label{loesung}
\mathbf{G}_{\mathbf{k}\sigma} =
\hbar\;(\mathbf{E}-\mathbf{H}_{\mathbf{k}\sigma})^{-1} \end{equation}

\subsection{Solution with Electric Field} 
\label{sec:feldloesung} 
The Hamiltonian of the electric field (\ref{feldham}) is a
single-particle-operator: \begin{equation}H_{V} =\sum_{ij\sigma \atop
\alpha\beta} \;M_{ij}^{\alpha\beta}\;c_{i\alpha\sigma}^+c_{j\beta\sigma} = -u\;
\sum_{i\alpha\sigma} \;\alpha\,n_{i\alpha\sigma} \;. \end{equation} Expressed
in terms of second quantization, i.e. by means of the electron  creation- and
annihilation-operators the electric-field-operator $H_{V}$ is therefore of the
same structure as  the operator of the kinetic energy of electrons $H_{s}$.
They can be summed up:  \begin{eqnarray} H = H_{s}+H_{V}+H_{sf}
&=&\sum_{ij\sigma \atop \alpha\beta} \;\widetilde{T}_{ij}^{\alpha\beta}
c_{i\alpha\sigma}^+c_{j\beta\sigma} + H_{sf}\\
T_{ij}^{\alpha\beta}\longrightarrow\widetilde{T}_{ij}^{\alpha\beta}
\;\;\;\;&\stackrel{\rm def}{=}&
T_{ij}^{\alpha\beta}\;\;\;\;-\alpha\,u\;\delta_{ij}^{\alpha\beta}
\end{eqnarray} The renormalisation of the hopping matrix elements by the
external field is equivalent to a displacement of the Bloch-band centers of
gravity by a position dependent amount for each layer.

An external electric field thus renormalizes the layer dependent self-energy:
\begin{equation}\label{SiRe}
\Sigma_{\sigma}^{\alpha}\longrightarrow\widetilde{\Sigma}_{\sigma}^{\alpha}\stackrel{\rm
def}{=} \Sigma_{\sigma}^{\alpha}-\alpha\,u \quad. \end{equation} However, the
principal structure of the  many-body problem remains unaffected. Its solution
is given by \begin{equation}\label{gensol} \mathbf{G}_{\mathbf{k}\sigma} =
\hbar\;(\mathbf{E}-\mathbf{\widetilde{H}}_{\mathbf{k}\sigma})^{-1},
\end{equation} with $\mathbf{\widetilde{H}}_{\mathbf{k}\sigma}$ defined
analogously to (\ref{hk}) with the renormalized self-energy (\ref{SiRe}).

The matrix inversion in (\ref{gensol}) is done numerically for a sc-film with
layers parallel to the (100)-face.

\section{Results}
\label{sec:results}
As stated above, the self-energy for 
$\sigma=\uparrow$-electrons is trivial. This is due to the fact, that this
electron group cannot participate in the spin flip processes, so that its
interaction with the ferromagnetically saturated 4-f spins is restricted to
rigid energy shift  of the $\sigma=\uparrow$-conduction band.  For that reason
we now focus on the discussion of the $\sigma=\downarrow$-results. 

\subsection{Spectral Densities} 
With the Green functions given by
(\ref{gensol}) we can  calculate the  spectral densities
$S_{\mathbf{k}\sigma}^{\alpha\alpha}$.  They depend on the spin direction
$\sigma$ of the test-electron  as well as on the layer-index $\alpha$ and  the
in-plane wave-vector $\mathbf{k}$: \begin{equation}\label{sksial}
S_{\mathbf{k}\sigma}^{\alpha\alpha}(E)=-\frac{1}{\pi}\;\Im\{G_{\mathbf{k}\sigma}^{\alpha\alpha}(E+i0^+)\}\quad,
\end{equation} where $G_{\mathbf{k}\sigma}^{\alpha\alpha}$ is the element
$(\mathbf{G}_{\mathbf{k}\sigma})_{\alpha\alpha}$ of the Green function matrix
(\ref{gensol}).

Fig.\ \ref{fig4} shows the spectral density
$S_{\mathbf{k}\sigma\downarrow}^{3,3}$  for the center layer ($\alpha=3$) of a
five-layer-film with  and without (right/left column) external electric field 
for no, medium and strong (top/ middle/ bottom line) interaction between
4f-spins and conduction  band. The top left picture of Fig.\
\ref{fig4} shows accordingly the well known dispersion of the free
Bloch electron gas in the two-dimensional Brillouin zone. 

As the system consists of five layers one might expect five excitation
energies.  However, for symmetry reasons  two pairs of them are degenerated, we
see only three peaks at each $\mathbf{k}$-point.\\ By switching on an external
electric field perpendicular to the film, this degeneracy is lifted  and  five
different peaks can be observed  (top right picture of Fig.\ 
\ref{fig4}).

Middle and bottom line of Fig.\ \ref{fig4} show the situation for
finite values of the coupling strength $J$ between 4f-spins  and conduction
band electrons. In each of these pictures we find two main structures: an
energetically lower lying, broad and  dispersionless so called \emph{flip
band}  and  at higher energies a sharply peaked structure which corresponds to
the \emph{magnetic polaron} and shows a Bloch like dispersion.

The flip band originates from spin flip processes of the conduction band
electrons. Via the s-f interaction, the $\sigma=\downarrow$ conduction band
electron can emit a magnon, i.e. it causes a spin deviation in the
ferromagnetically saturated lattice of 4f-spins. As this process is of course
forbidden for $\sigma=\uparrow$ electrons, the spectral densities 
$S_{\mathbf{k}\sigma\uparrow}$ for $\sigma=\uparrow$ electrons are rather
trivial. They are not shown separately.

The sharply peaked structure corresponds to a quasiparticle with in most cases 
infinite lifetime: the magnetic polaron. This means, in analogy to the
"normal"  polaron, an electron, renormalized by a cloud of virtual excitations,
namely magnons. As this quasiparticle propagates freely through the crystal
lattice,.  the corresponding structures in the spectral densities show
therefore  a Bloch like dispersion.

In the strong coupling regime ($J>$0.2 eV) both structures are energetically
clearly separated, the polaron is represented by  $\delta$-peaks and has
therefore indeed infinite lifetime. This changes at weaker couplings: At
$J\approx0.2 eV$, i.e. realistic values for EuS, the polaron peak  touches the
flip band and becomes considerably broadened near to the $\Gamma$-point of the
two dimensional Brillouin zone. The lifetime of the polaron under these
circumstances is finite and of the order of  \mbox{1 $\hbar/eV \approx 10^{-15}
sec$.}

The external electric field (right column of Fig.\ \ref{fig4})
increases mainly the energetic distances between the layers and thereby
influences strongly the shape of all spectral densities. The polaronic peaks 
become more separated. As the flip band  reproduces roughly the shape of the
($\sigma=\uparrow$)-DOS, the flip band changes with  the 
($\sigma=\uparrow$)-DOS under the influence of the electric field, too. Because
of (\ref{w=S1}) one should expect a similarly significant field induced change
of the transition- and spin flip- probabilities. This will be investigated in
the following  sub-section.

\subsection{Probabilities} 
\label{subsec:prob} 
We apply (\ref{w=S1}) on the
typical $\sigma=\downarrow$-spectral density of the s-f model. As calculated
(\ref{sksial}, \ref{gensol}) and shown above  Fig.\ \ref{fig4}  it
consists at each $\mathbf{k}$-point of the  broad flip band, written now as
$f(E)$, a continuous function of energy without  singularities, and the
$\delta$-like polaron band $\hbar\delta(E-E_0)$: \begin{equation}\label{Ssf}
S(E)=\alpha_1\;f(E)+\alpha_2\;\hbar\delta(E-E_0)\;. \end{equation} The
$\alpha_n$  are the spectral weights ($\alpha_1+\alpha_2=1$). 

From (\ref{Ssf}) one finds with (\ref{w=S1})  \begin{equation}\label{w=S1b}
\bar{p}(t)=\big|\alpha_1\;\widetilde{f}(t)+\alpha_2\;e^{-\frac{i}{\hbar}E_0
t}\big|^2 \end{equation} because of the linearity of Fourier transformation.
With Parsevals theorem \begin{equation}\label{parseval}
\int_{-\infty}^{+\infty}\big| f(E)\big|^2\;dE=\int_{-\infty}^{+\infty}\big|
\widetilde{f}(t)\big|^2\;dt \end{equation} we conclude from the normalization
of $f(E)$ that $|\widetilde{f}(t)|^2$ must vanish quicker than $1/t$ for
$t\rightarrow\infty$. 

For a typical spectral density of our model we therefore find from (\ref{Ssf})
and  (\ref{w=S1}): \begin{equation}\label{winfty}
\bar{p}(t)\stackrel{t\,\rightarrow\,\infty}{-\!\!\!-\!\!\!-\!\!\!-\!\!\!-\!\!\!\longrightarrow}\big|
\;\alpha_2\; \big|^2 \end{equation} This interesting result shows, that the
non-flip probability for a ($\sigma=\downarrow$)-electron is completely
determined by the spectral weight of the polaron peak in the spectral density
and therefore strongly dependent on position in the Brillouin zone, coupling
strength and temperature.

We ask now for the probability of finding an
electron independent of its spin direction  in a  layer with index $\beta$ at
time $t$ after we prepared it at  time $t=0$ in layer $\alpha$. Generalizing
(\ref{w=S1}) we find: \begin{equation}\label{w=S2}
\bar{p}^{\alpha\beta}_{\mathbf{k}}(t)=4\pi^2 \;\big|
S_{\mathbf{k}}^{\alpha\beta}(t)\big|^2 \;. \end{equation} We will illustrate
(\ref{w=S2}) by investigating it analytically for a two layer system of free
electrons (i.e. $J=0$) with applied electric field.  The spectral densities
have a double peak structure: \begin{equation}\label{spekmat1}
S_{\mathbf{k}}(E)=\hbar\;\big[\alpha_1\;\delta(E-E_1({\mathbf{k}}))+\alpha_2\;
\delta(E-E_2(\mathbf{k}))\big], \end{equation} where the $\alpha_n$ are again
the spectral weights of the $\delta$-peaks  ($\alpha_1+\alpha_2=1$) and $E_n$
is the $n$-th excitation energy of the system.   $E_n$ is given as the $n$-th
pole of the Green functions (\ref{gensol}): \begin{eqnarray} \label{poles} {\rm
det} (\mathbf{E}-\mathbf{H}_{\mathbf{k}})\Big|_{E=E_{n}}&\stackrel{!}{=}&0
\nonumber \\ \label{anrzwei}
E_{1/2}(\mathbf{k})&=&\frac{1}{2}\Big(2\epsilon(\mathbf{k})-u\pm\Delta\Big)\nonumber\\
\Delta&=&\sqrt{u^2+4t^2} \end{eqnarray} $\Delta=E_1-E_2$ is a measure for the
energetic separation of the layers and depends on the interlayer hopping $t$
set to $0.1 eV$ and the interlayer- potential-difference $u$ induced by the
external electric field (\ref{feldham}). 

Now a simple  calculation from (\ref{w=S2}) using (\ref{spekmat1}) yields 
\begin{eqnarray} 
\label{w12} 
\bar{p}^{1\,2}(t)&=&\bar{p}^{2\,1}(t)
=4\left(\alpha^{1\,2}\right)^2 \sin^2\left(\frac{t\Delta}{2\hbar}\right)\;,\\ 
\label{w11} 
\bar{p}^{1\,1}(t)&=& \bar{p}^{2\,2}(t)=1-\bar{p}^{1\,2}(t)\;,
\end{eqnarray} 
the  probability of finding an electron in the second layer
($\bar{p}^{12}$) or in the first layer ($\bar{p}^{11}$) at time $t$ after we
prepared it at time $t=0$ in the first layer of our two layer film. Both
probabilities are completely determined by $\Delta$, the energetic separation
of the layers and therefore not $\mathbf{k}$-dependent. Increasing the external
electric field $u$, we observe a growing confinement of the electron in the
layer it had been prepared in Fig.\  \ref{fig6}.
This behavior is well known as
Wannier-Stark-localization\cite{Wann60,Kane59,Shoc72}.

Finally we want to determine the probability of
finding an electron at time $t$  in layer $\beta$ with reversed spin $-\sigma$
after it had been prepared at  time $t=0$ in layer $\alpha$ with opposite spin
$\sigma$. These spin flip probabilities $p_{\mathbf{k}\sigma}^{\alpha\beta}(t)$
are obviously given by the overlap of the Bloch states
$|\psi_{\mathbf{k}\alpha\sigma}(0)\; \rangle =
c_{\mathbf{k}\alpha\sigma}^+|0\;\rangle$ with states like
$|\phi_{\mathbf{k}\mathbf{q}-\sigma}^{\beta\gamma}(t)\;\rangle=|S_{\mathbf{q}\gamma}^{\sigma}
c_{(\mathbf{k-q}),\beta,-\sigma}(t)\;|0\rangle$.

The spin flip process of the conduction band electron   results always from a
magnon emission, i.e. it is connected to a spin deviation in the lattice of the
magnetic 4-f moments. Let us assume this magnon had been emitted in layer
$\gamma$ and carries away the momentum $\mathbf{q}$.  In order to get the total
spin flip probability $p_{\mathbf{k}\sigma}^{\alpha\beta}(t)$ we have to sum
over all possibilities of emitting such a $(\mathbf{q},\gamma)$-magnon:
\begin{equation}\label{defwhat}
p_{\mathbf{k}\sigma}^{\alpha\beta}(t)=\sum_{\gamma}\sum_{\mathbf{q}}\;\big|\langle 
\phi_{\mathbf{k}\mathbf{q}-\sigma}^{\beta\gamma}(t)|\psi_{\mathbf{k}\alpha\sigma}(0)\;\rangle\big|^2.
\end{equation}

The flip probabilities $p_{\mathbf{k}\sigma}^{\alpha\beta}(t)$ are related to
the  spectral density 
$\widehat{S}_{\mathbf{k}\mathbf{q}\sigma}^{\gamma\beta\alpha}$ of the spin flip
Green function  (\ref{flipfkt})
\mbox{$F_{\mathbf{k}\mathbf{q}\sigma}^{\gamma\beta\alpha}=\left\langle\!\left\langle
S^{-\sigma}_{-\mathbf{q} \gamma} \; c_{(\mathbf{k-q}),\beta,-\sigma}\, ;
\,c_{\mathbf{k}\alpha\sigma}^+ \right\rangle\!\right\rangle$}.  

Straight forward calculation similar to those in section  
(\ref{subsec:prob_green}) show: 
\begin{equation}\label{flipwl}
p_{\mathbf{k}\downarrow}^{\alpha\beta}(t)=4\pi^2\;\sum_{\gamma\mathbf{q}}\;
\big|\widehat{S}_{\mathbf{k}\mathbf{q}\downarrow}^{\gamma\beta\alpha}(t)
\big|^2 
\end{equation} 
The corresponding probability
$p_{\mathbf{k}\uparrow}^{\alpha\beta}(t)$ vanishes, since we considered here
the case of a ferromagnetically saturated 4-f lattice, which  cannot be aligned
any further and thus interdicts spin flip processes for
($\sigma=\uparrow$)-electrons.

In a last step we have now to determine the general spin flip function
$F_{\mathbf{k}\mathbf{q}\downarrow}^{ \gamma\beta\alpha}$ which is in the usual
way connected to its spectral density:
\begin{equation}\widehat{S}_{\mathbf{k}\mathbf{q}\downarrow}
^{\gamma\beta\alpha}(E)=-\frac{1}{\pi}\Im\{F_{\mathbf{k}\mathbf{q}\downarrow}^{\gamma\beta\alpha}
(E)\}\;. \end{equation} After Fourier transformation we will get  
$\widehat{S}_{\mathbf{k} \mathbf{q}\downarrow}^{\gamma\beta\alpha}(t)$.

The general spin flip function can  be obtained from the hierarchy of the
equation of motion. The calculation is comparable to the one in section
(\ref{sec:sol}) and  yields finally: \begin{equation}\label{flipgl1}
F^{\alpha\gamma\beta}_{\mathbf{kq}\downarrow}(E)=
X^{\alpha}(E)\;G_{\mathbf{k}\downarrow}^{\alpha\beta}(E)\;G_{\mathbf{k-q}
\uparrow}^{\gamma\alpha}(E)
\;, \end{equation} with  \begin{equation}\label{defxal}
X^{\alpha}(E)\;\stackrel{\rm def}{=} \;\frac{J \hbar S}{\sqrt{N_s}}
\left(\frac{2}{J\hbar B_{\alpha}(E)-2}\right). \end{equation} The equations
(\ref{hk}-\ref{loesung}) and their generalized form which includes the external
electric field (\ref{SiRe}),(\ref{gensol}) completely determine  the solution.
(\ref{flipwl}) and (\ref{flipgl1}) give us  the  probabilities for spin flip
processes we were looking for. 


The  spin flip
- and non flip- probabilities derived so far were evaluated numerically for a 
two-layer sc-film with layers parallel to the (100)-face. The results for
different values of s-f coupling and strength of the external electric field
are shown in the Figures \ref{fig7},\ref{fig8} and \ref{fig9}.

It should be stressed, that the shown  non-flip- and flip-probabilities were
obtained in separate calculations, based on completely different many-body
entities (one-electron Green function and spin flip function, respectively). 
However, our understanding of those complementary probabilities demands that
they add up to 1 for all times and all vectors $\mathbf{k}$ (particle
conservation). As the figures show, this is fulfilled in each case (bottom
right pictures).

Comparing the Figures \ref{fig7}-\ref{fig10} we can summarize:
\begin{enumerate} \item	The spin-totalized probabilities (bottom lines) do not
show a  significant dependence on the strength of the s-f coupling $J$  between
conduction band electrons and localized 4-f moments.  Accordingly they are
equal to those with no coupling $J\equiv0$  which had been determined
analytically in subsection  (\ref{subsec:prob}) and are completely independent
of the  intra-layer momentum $\mathbf{k}$.    \item	The total spin flip - and
non flip- probabilities (right columns) do not show any dependence on the
external electric field $u$. Accordingly they are equal to those without field
$u\equiv0$. \item	Calculations for films with various numbers of layers show:
The total spin flip - and non flip- probabilities are also independent from the
total number of layers in the film. Consequently they are the same as  those
which are calculated for a mono-layer.  \end{enumerate}

These observations simplify the practical evaluation of spin flip probabilities
for the spin filter experiment essentially and allow to determine the
polarization of the field-emitted electrons  quite generally.

\subsection{Polarization}

According to (\ref{p=w}) the polarization is completely determined by the
effective spin flip ratio $p_{eff}=p_{\downarrow}-p_{\uparrow}$, where the flip
probability for ($\sigma=\uparrow$)-electrons $p_{\uparrow}$ vanishes
identically and the (time dependent)  flip probability for
$\sigma=\downarrow$-electrons $p_{\downarrow}$ is given by (\ref{flipwl}) as
pointed out  in the previous subsection. 

Using the observations (2) and (3) concerning
$p_{\mathbf{k}\uparrow}^{\alpha\beta}(t)$  made in the previous subsection the
calculation of the polarization simplifies considerably: The dynamics of the
spin flip processes turned out  to  depend neither on the strength of the
external electric field nor on the number  of layers of the spin filter. It is
completely determined by the strength of the s-f interaction $J$. 

Practically calculating the polarization therefore simply demands to evaluate
the spin flip probabilities for a mono-layer-film without electric field
$p_{\mathbf{k}\downarrow}$or, even simpler, its complementary quantity:  the
non flip probability $\bar{p}_{\mathbf{k}\downarrow}$ as 
\begin{equation}p_{\mathbf{k}\downarrow}+\bar{p}_{\mathbf{k}\downarrow}\equiv1\;
. \end{equation}

In the previous subsection we showed, that after a sufficient long period of
time  the non flip probability   $\bar{p}_{\mathbf{k}\downarrow}$ is given by
the square of the spectral weight of the polaron. This ''critical'' time is
characterized by the width of the scattering peak:
\mbox{$t\gtrapprox1\hbar/eV=6.6*10^{-16} sec$} and thus sufficiently  smaller
than typical amount of time an electron spends in the spin filter.  We can
therefore apply equation (\ref{winfty}) and obtain finally:
\begin{equation}\label{pk}
P_{\mathbf{k}}=p_{\mathbf{k}\downarrow}=1-\big|\alpha_2(\mathbf{k})\big|^2\;.
\end{equation} The degree of polarization is completely  given in terms of the
spectral weight of the magnetic polaron peak $\alpha_2(\mathbf{k})$ and thus
strongly dependent on the  s-f coupling as well as the electrons intra-layer
momentum $\mathbf{k}$.  $\alpha_2(\mathbf{k})$ is obtained by integration of
the spectral density  (\ref{sksial}).
Fig.\ \ref{fig11} shows the degree of polarization  $P_{\mathbf{k}}$ calculated 
according to (\ref{pk})  for different s-f interactions. 

It is interesting to see that for unrealistic small values of the coupling
($J\approx0.08eV$), one really gets nearly 100\% of polarization. This is due
to the finite life time of the magnetic polarons under  these circumstances:
Polaronic and scattering peak touch each other near to the origin of the
two-dimensional Brillouin zone, the polaronic gets broadened  and the
$\sigma=\downarrow$-electrons ''decay''  into ($\sigma=\uparrow$)-states. 

However, for realistic values of $J$ ($J^{EuS}\approx0.2 eV$) the polarization 
lies for all  $\mathbf{k}$ clearly below 100\% (Fig.\ \ref{fig11}, right).

The spin filter experiment  allows only for a $\mathbf{k}$-averaged measurement
of the degree of polarization $P$. Out of all electrons with given total energy
those electrons with maximum energy perpendicular to the tunneling barrier will
be transmitted most likely. That are just those electrons with minimum
intra-layer momentum $\mathbf{k}$, i.e. the electrons close to the
$\Gamma$-point of the two-dimensional Brillouin zone. According to our
calculation (Fig.\ \ref{fig11}, right) this would yield  $P\approx80\%$, which
accounts much better for the experimental results then former mean field
results do.

To exploit equation  (\ref{pk}) fully, $\mathbf{k}$-resolved experiments as
for  instance Spin Polarized Low Energy Electron Diffraction (SPLEED) or  Spin
Polarized Electron Energy Loss Spectroscopy (SPEELS) on an europium 
chalcogenide surface are suggested.  Because of (\ref{pk}) one would expect  a 
significant angle- i.e. $\mathbf{k}$- dependence of the degree of polarization
of  the scattered electrons (see Fig.\ \ref{fig11}).

\section{Concluding Remarks} 
\label{sec:conc}

Aim of the presented paper was a new interpretation of the spin filter
experiments done on Europium chalcogenides using recent results from
many-body-theory on the s-f model. This was  achieved by expressing the  degree
of polarization of the field emitted electrons  in terms of spin flip
probabilities which had been  determined  in the framework  of an exactly
solvable many-body model of the experimental situation. This model used the s-f
model in reduced dimensions (film geometry), including an additional term
taking into account the  strong external electrical field.

Discussing the spectral densities of the one-particle Green function,  the
influence of the external electrical field on electronic behavior was seen 
e.g. as Wannier-Stark-ladder.

Several physically relevant probabilities as layer resolved probability
densities, flip and non flip probabilities had been derived and discussed,
showing for instance the Wannier-Stark-localization due to the electrical field
and the strong dependence of the spin flip probabilities on the intra-layer
momentum and the s-f coupling between ferromagnetically ordered 4-f spins and
conduction band electrons of the spin filter.

The degree of polarization of the field emitted electrons evaluated in our
model turned out to be well below 100\% for all temperatures which is  in good
agreement with the experiments and  a considerable progress  with respect to
mean field results of former works.

\acknowledgments
This work has been supported by Studienstiftung des
Deutschen Volkes.


\begin{figure}
\caption{Temperature dependent quasiparticle bandstructure of the s-f model 
in Mean-Field-Approximation (schematically)} 
\label{fig1} 
\end{figure}

\begin{figure}
\caption{The temperature dependent  degree of polarization for field emitted
electrons in a spin filter experiment. The dotted line is the prediction of the
often used mean field solution for the s-f model. Data were taken from Kisker
et al.\cite{KiBM78}, the point   (T=9K, $P$=85 \%) was taken from 
\cite{BKiM77}.}
\label{fig2} 
\end{figure} 

\begin{figure}
\caption{Decomposition of a linear potential into a step-like ($V_1$) and saw
tooth-like part. The points mark the field displaced centers of gravity of the
conduction band in the layers. $\alpha$ is the layer index.} 
\label{fig3} 
\end{figure}

\begin{figure}
\caption{Density plot of the spectral densities
$S_{\mathbf{k}\sigma\downarrow}$ of the middle layer of a 
five layer spin filter film with  and without
(right/left column) external electric field for several values  of coupling
strength $J$  between 4f-spins and conduction band} 
\label{fig4} 
\end{figure}

\begin{figure}
\caption{The non-flip probability for electrons in the spin filter is given  by
the square  of the time dependent spectral density of the one-electron  Green
function. Shown are typical shapes of the $\sigma=da$-Spectral densities of the
spin filter model:  energy-dependent (left) and time-dependent (middle). and
the corresponding time-dependent non-flip probability (right).} 
\label{fig5} 
\end{figure}

\begin{figure}
\caption{Time dependent probabilities $\bar{p}^{1\,1}(t)$ (starting at p=1)
and  $\bar{p}^{1\,2}(t)$ (starting at p=0) of finding an electron in the first
or second layer at time $t$ for different values of the external field ($u$). 
Time is given in $\hbar/eV\;=\;6.6*10^{-16} sec$. Straight lines are time
average.} 
\label{fig6} 
\end{figure} 

\begin{figure}
\caption{Spin flip - and non flip- probabilities for a two-layer film with weak
s-f coupling ($J=0.08 eV$) and without electric field. At $t=0$ a
$\sigma=\downarrow$-electron had been prepared in layer one. The first line of
pictures shows the time-dependent  probability  of still being in a
$\sigma=\downarrow$-state (in first layer, second layer,  somewhere in the
system). Second line of pictures shows the  probability  of being flipped (in
first layer, second layer,  somewhere in the system). Bottom line is the sum
over both spin directions and  gives thus the spin independent probability of
being in the (first layer, second layer, somewhere in the system). Time is
given in $\hbar/eV=6.6*10^{-16} sec$; the interlayer momentum $\mathbf{k}$ runs
from (0,0) to ($2\pi,2\pi$) through the two-dimensional Brillouin-Zone.}
\label{fig7} 
\end{figure}

\begin{figure}
\caption{Same as Fig.\  \ref{fig7}, but with external electric field  ($u=0.4$ 
eV). The total spin flip - and non flip- probabilities (right column) are
-compared to Fig.\  \ref{fig7}- unchanged, although the individual probabilities
changed rather drastically. The spin-totalized probabilities (bottom line) show
the Wannier-Stark-Localization discussed in section  (\ref{subsec:prob}): the
electron is confined to the layer it had been prepared  in.} 
\label{fig8}
\end{figure}

\begin{figure}
\caption{Same as
Fig.\  \ref{fig7}, but with intermediate s-f coupling ($J=0.2$ eV) and 
without electric field. The total spin flip - and non flip- probabilities
(right column) show -compared to Fig.\  \ref{fig7}-  a strong dependence
from the s-f coupling. The spin-totalized probabilities (bottom line) however
are barely changed.} 
\label{fig9} 
\end{figure}

\begin{figure}
\caption{Non-flip-
and flip-probabilities for a mono-layer spin filter with weak s-f coupling
($J=0.08$ eV). Time is given in $\hbar/eV=6.6*10^{-16} sec$; the interlayer
momentum $\mathbf{k}$ runs from (0,0) to ($2\pi,2\pi$) through the
two-dimensional Brillouin-Zone. The right picture verifies the particle
conservation.} 
\label{fig10} 
\end{figure}

\begin{figure}
\caption{Degree of polarization  $P_{\mathbf{k}}$ according to (\ref{pk})  for
weak  ($J=0.08$ eV, left) and intermediate    ($J=0.2$ eV, right) s-f
interaction.  The right picture shows the situation for realistic values of $J$
in an spin filter experiment: $J^{EuS}\approx0.2$ eV following Wachter et
al.\cite{Wac79}.}
\label{fig11} 
\end{figure}

\end{document}